# Label-free detection of miRNAs: role of probe design and bioassay configuration in Surface Enhanced Raman Scattering based biosensors

*Chiara Novara[1], Daniel Montesi[1], Sofia Bertone[1], Niccolò Paccotti[1], Francesco Geobaldo[1], Marwan Channab[1,2], Angelo Angelini[2], Paola Rivolo[1], Fabrizio Giorgis[1], Alessandro Chiadò[1]*

[1]Department of Applied Science and Technology, Politecnico di Torino, Corso Duca degli Abruzzi, 24, 10129 Turin, Italy.

[2]Advanced Materials and Life Sciences, Istituto Nazionale di Ricerca Metrologica (INRiM), Strada delle Cacce 91, Turin 10135, Italy.

**ABSTRACT.** Accurate design of labelled oligo probes for the detection of miRNA biomarkers by Surface Enhanced Raman Scattering (SERS) may improve the exploitation of the plasmonic enhancement. This work, thus, critically investigates the role of probe labelling configuration on the performance of SERS based bioassays for miRNA quantitation. To this aim, highly efficient SERS substrates based on Ag-decorated porous silicon/PDMS membranes are functionalized according to two different bioassays relying on a one-step or two-step hybridization of the target miRNA with oligonucleotide probes. The detection configuration is varied, exploring the impact of the position of the Raman reporter along the oligo sequence and of the reporter identity on bioassay sensitivity. In the high miRNA concentration regime(100-10 nM), a significantly increased SERS intensity is detected when the reporters are located closer to the plasmonic surface compared to farther probe labelling positions. Counterintuitively, a levelling-off of the SERS intensity from the two configuration is instead recorded at low miRNA concentration. Such effect is explained by an increased contribution of Raman hot spot to the whole SERS signal, as confirmed by simulations of the electric near field for a simplified model of the Ag nanostructures. The beneficial effect of reducing the reporter-to-surface distance is however partially retained for the two-step hybridization assay thanks to the less sterically hindered environment in which the second hybridization occurs. The study thus demonstrates that the limit of detection of the assay can be lowered by tuning the probe labelling position, but sheds at the same time light on the complexity of the bionanointerfaces in SERS and on the multiple factors affecting bioassay sensitivity.

**Introduction**

Surface Enhanced Raman Scattering (SERS)-based sensors provide excellent results in the detection of biomolecules aimed to the easy and rapid diagnostic and follow-up monitoring of biomarkers.[1] Besides the development of tailored SERS active platforms, thanks to the improved knowledge on the SERS enhancement mechanisms and to the increased availability of synthetic procedures for Raman hot-spot engineering, the relevance of the functionalization strategy of the plasmonic nanostructures to enhance the sensitivity and the specificity of the detection has been fully recognized.[2] Actually, the use of specific capture probes anchored to the nanoparticles surface is a widely adopted approach for biomarker detection, especially moving towards their analysis in complex mixtures such as biofluids or food extracts.[3,4] In such a framework, the direct detection of the analyte requires the processing of the complex spectral profiles by chemometric/machine learning tools[5] and is often challenging if very low concentrations need to be addressed. Moreover, the use of multiple probes in sandwich or competitive assays allowed introducing Raman reporters preserving a label-free approach.[6,7] The target biomolecule could thus be detected without the need of chemical modifications, but with an improved sensitivity thanks to the very high Raman cross section of the reporter, possibly excited in electronic resonance Raman conditions.

Among the multitude of interesting applications in biomedicine, microRNA (miRNAs) detection strategies are being deeply investigated.[8] Such short non-coding RNA strands, involved in the regulation of several biological processes through the control of gene expression, are widely studied as markers of cancer disease.[9] Indeed, the alteration of the concentrations of specific miRNA sequences in tissues and body fluids has been related to the onset of cancer pathologies[10]. Thus, miRNA profiling has a great potential for early tumour diagnosis. In such framework, SERS sensors provide good sensitivity, specificity, wide dynamic ranges and multiplexing capability at lower cost and in shorter time with respect to traditional nucleic acid analytical techniques[6,11]. Complementary DNA probes have been used to functionalize the plasmonic nanostructures. Recently, new protocols including in situ amplification steps relying both on the use of DNA cleaving or replicating enzymes or on chip PCR have appeared[12] Other works reported enzyme-free approaches[13], even based on HCR (Hybridization Chain Reaction) amplification[11] or the use of SERS tags able to bind the captured miRNA sequence[7,14]. In such cases, the reported limits of detection are lowered down to the pM-fM concentration range, suitable for miRNA profiling in blood plasma[8], but the procedures are significantly complicated and featured by increased costs. Actually, the use of Raman reporters in miRNA detection assays is widespread and different probes were designed bearing the reporter in different position with respect to the plasmonic surface, such as in the case of ON-OFF assays based on the plasmonic enhancement variation upon a change of the surface to molecule distance in the opening of loop probes[15]; however, a critical comparison between different configurations was not considered. Indeed, an improvement of the analyte detection limits is envisaged if the 'plasmonic surface- Raman reporter' distance is reduced, due to the strong increase of SERS electromagnetic enhancement when the fluorophore-NPs spacing is decreased, together with a potential quenching of the deleterious photoluminescence yield of the reporter.

In this work, we tune the performance of a SERS based assay for the label-free detection of miR-222 through a detailed study of different detection configurations in which the distance of the reporter from the nanostructured surface is varied. In order to widen the range of achievable spacings both a one-step and a two-step hybridization assays are tested using Ag-decorated porous silicon nanostructures as SERS active platform. The behaviour of each configuration versus miR-222 concentration is critically analysed and some unexpected trends, such as a reduced distance dependence of the SERS amplification at low target concentration, are understood in terms of the plasmonic properties of the SERS substrate and probe-target interaction at the silver surface.

## Experimental Procedures

### Chemicals and materials

DL-Dithiothreitol (98.0% DTT), acetic acid (99.0%), bovine serum albumin (BSA, IgG free), polyethylene glycol sorbitan monolaurate (Tween), sodium acetate (99.0%), sodium chloride (99.5%), tris(hydroxymethyl)aminomethane (TRIS), ethanol (>99.8%), ethylenediaminetetraacetic acid (EDTA, 99.4%), saline sodium citrate (SSC, sterile stock 20x, 300 mM trisodium citrate, 3 M NaCl), sodium dodecyl sulfate solution (BioUltra, 10%, SDS), 3,30 ,5,50 -tetramethylbenzidine (TMB), silver nitrate, heat-inactivated fetal bovine serum. All of the DNA and RNA oligos were from Sigma Aldrich, Milan, IT, and from Integrated DNA technologies, Leuven, BE. HF was from Carlo Erba, Milan, IT. The PDMS pre-polymer and curing agent (Sylgard 184) were from Dow Corning and the silicon wafers from Sil'tronix Silicon

Technologies, Archamps, France. The Illustra MicroSpin G-25 columns were from GE Healthcare (Fisher Scientific, Illkirk, FR). The water used during each step was Milli-Q™ dispensed from a DirectQ-3UV (Merck-Millipore, Milan, IT) and sterilized.

**Ag-Porous silicon-PDMS SERS platforms synthesis.**

Ag-porous silicon-polydimethylsiloxane membranes (Ag-PSD) were prepared according to a previously reported procedure[16]. Briefly, boron-doped Si wafers (34 mΩ cm resistivity) were electrochemically etched in 20 : 20 : 60 - $HF/H_2O/CH_3CH_2OH$ solutions, with a current density of 125 mA/cm$^2$ for 30 s. The obtained mesoporous membranes, featuring a 2 μm thickness and 80% porosity, were then partially lifted off by means of a second electrochemical attack with a 2% HF nearly ethanolic solution, at a current density of 4 mA/cm$^2$ for 95 s, and transferred on partially crosslinked polydimethylsiloxane (PDMS) membranes obtained by mixing in a 10:1 weight ratio the oligomer and the curing agent, respectively. After a refresh in 5% HF aqueous solution and thorough rinse in ethanol, the obtained substrate was dried and finally dipped in 10$^{-2}$ M silver nitrate aqueous solution, supplemented with HF up to a concentration of 0.0025%. Indeed, by this step, the silver nanoparticles (NPs) were *in situ* synthesized, by exploiting the reducing properties of Si-H$_x$ (x=1-3) groups covering the surface of freshly etched porous silicon (pSi). After an incubation time of 4 s, the Ag-PSD substrates were rinsed in dH$_2$O and gently dried under N$_2$ flux.

**Morphological characterization of the SERS substrate**

Field Emission Scanning Electron Microscopy (FESEM) was performed to assess the morphology of the synthesized silver nanostructures. Due to the poor conduction by the PDMS support, samples were covered by a copper grid to allow electron discharging. Images were acquired by a Zeiss Supra 40 FESEM using 20 µm aperture and 3 keV primary beam with secondary electrons analysis.

**Plasmonic properties simulation**

A simplified system of the SERS active substrates consisting in Ag hemispheres dimers on pSi was modelled by 3D Finite Element Method (FEM). At this aim the *Electromagnetic Waves, Frequency Domain* module of the COMSOL Multiphysics software (version 5.1) was exploited. The excitation laser was simulated through a background electric field, $E_{bk}$ = 1 V/m with a polarization plane along the axis of the two AgNPs at different wavelength, to study the coupling of their surface plasmons in resonant conditions. The electric near-field intensity in stationary conditions was calculated for representative NPs diameter-gap size combinations and analysed inside and outside the inter-particle gap, in order to get insights concerning their impact on the differences in the miRNAs detection highlighted in the subsequent experiments.

**One-step and two-step hybridization assays**

After a BSA blocking step aimed to reduce non-specific binding, the prepared Ag-PSD solid substrates were functionalized for miR-222 detection according to two previously optimized protocols[6]. In detail, as for the one-step hybridization assay, a 5' thiol modified DNA probe (probe-222, 5'-C6SH-ACCCAGTAGCCAGATGTAGCT-3') complementary to miR-222 sequence (5'AGCUACAUCUGGCUACUGGGU-3') was immobilized on the SERS substrate surface overnight in Tris-EDTA 1 M NaCl (TE NaCl pH 7.5) buffer at 4 µM concentration. A shorter sequence, consisting of the first 11 nucleotides of probe-222 (half1, 5'-C6SH-ACCCAGTAGC-3') was instead used as a capture probe for the two-step hybridization assay. The same immobilization conditions were employed, except for the concentration that was lowered to 2 µM. miR-222 hybridization was then performed in SSC 5x supplemented with 0.05% tween20. A Raman reporter-labelled miRNA (miRNA-R) was employed during the one-step hybridization assay. The unlabelled miRNA was instead used in the two-step assay, since a sensitive detection was allowed in the second hybridization step performed in SSC 5x supplemented with 0.05% tween20 and 1% BSA using a labelled half2 probe (5'-CAGATGTAGCT) complementary to the unpaired part of the miRNA sequence. After each functionalization step three washings of 5 minutes were performed in appropriate buffers, as detailed elsewhere[7].

**Detection configurations**

Four configurations were studied. In detail, miR-222 was modified with a Raman reporter (R) at either the 5' or 3' terminus of the sequence in order to achieve significantly different reporter-to-surface distances. Similarly, the half2 probe was labelled either at 3' or 5' end. Two widely employed Raman reporters, Cyanine 3 (Cy3) and Rhodamine 6G (R6G), were used. UCSF Chimera package (1.16) was employed to calculate the dimensions of the probe to estimate the reporter-to-surface distances for each configuration and to prepare the graphical images[17]. The performance of the configurations was investigated by the analysis of different miR-222 concentrations (100 nM, 50 nM, 25 nM, 10 nM, 5 nM, 2.5 nM, 1 nM, 0.5 nM, 0.25 nM, 0.1 nM) in buffer solutions, for the selected Raman reporters (8 different combinations as a whole). For each configuration the experiments were repeated at least three times.

**SERS analyses**

All the SERS analyses were performed using a Renishaw InVia Raman microscope, equipped with a 514.5 nm laser line. The measurements were collected in backscattering configuration with a 100X long working distance objective (NA 0.75) and the laser power (100 mW) reduced to the 0.05 % through neutral density filters. A 5%-defocalization of the laser spot was applied to keep the power density as low as possible to avoid any oligonucleotide degradation. Each sample was analysed by recording a map of 100 spectra distributed over a grid of ca. 45 μm x 45 μm with a step size of 5 μm. The total acquisition time was 8 s for each spectrum, divided in 4 accumulations. The spectra were then analysed by a combination of the HyperSpec R package[18] and the Renishaw software WiRe 3.4. In detail, the baseline subtracted data set and the average spectrum of each map were calculated on R, while a deconvolution of selected regions of the spectra was carried out on WiRe 3.4 to obtain the integrated area value of specific Raman reporter bands for each recorded spectrum. The intra-substrate variability was expressed in terms of Relative Standard Deviation (RSD).

The obtained integrated areas were used to build calibration curves (vibrational band area vs. miR-222 concentration) for the different configuration-reporter combinations. Linear regression was performed and the Limit of Detection (LOD) of the different methods were calculated as the concentration corresponding to the average value of control samples increased by three standard deviations, as described in the literature.[19]

## 3. Results and Discussion

Ag-decorated porous silicon-PDMS substrates (Ag-PSD) were previously demonstrated as sensitive and reliable SERS platforms featured by a homogenous distribution of densely packed silver nanostructures[16] (a representative FESEM micrograph is reported in Figure S1). In this study, Ag-PSD were thus selected for the comparison of the different detection configurations in the framework of miRNA analysis. Figure 1 depicts the investigated configurations, showing the last hybridization step of a one-step and a two-step hybridization assay.

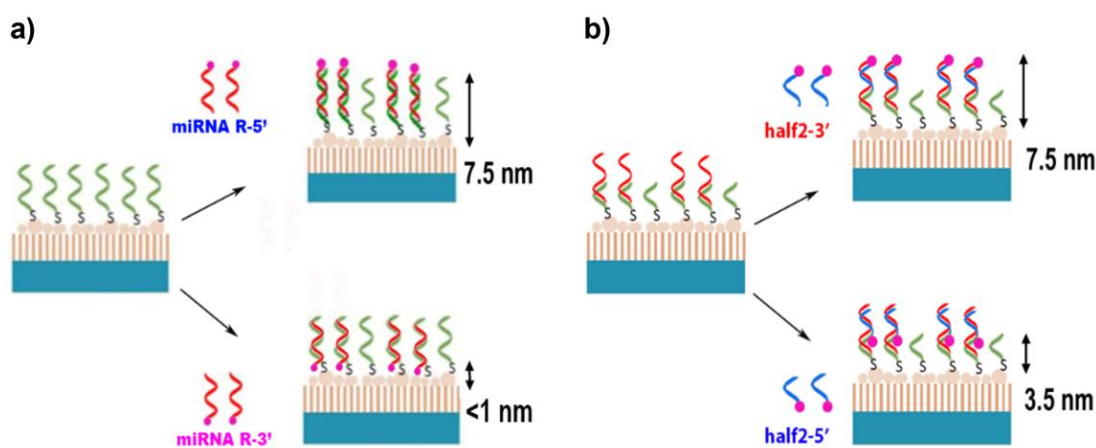

Figure 1. Scheme of the four detection configurations a) one-step assay: a fully complementary probe immobilized on the AgNPs binds the miRNA, which is labelled with a Raman reporter either at the 5' or 3' terminus; b) two-step assay: a first half-complementary probe (half1) immobilized on the AgNPs binds the unlabelled miRNA and the reporter is attached either at the 5' or 3' terminus of the half2 probe. The estimated reporter-to-surface distance for each configuration is highlighted. Note that NPs and oligos size are not to scale.

Concerning the one-step assay, the miRNA is recognized by a fully complementary DNA probe immobilized at surface of Ag NPs via a thiol group and two configurations are possible: one with the miRNA modified with a Raman reporter at the 5' terminus (miRNA R-5'), the other with the Raman reporter at the 3' terminus (miRNA R-3') (Figure 1a), namely bearing the reporter at the farthest end of the sequence or close to the NPs surface. The two-step assay, instead, has a great potential for a label-free detection in real samples, since the target miRNA does not require chemical modification, as its specific recognition and detection are accomplished by two probes, each complementary to half of its sequence: the thiolated "half1" probe is anchored to the NPs surface and captures the miRNA from the sample, while the sensitive detection occurs thanks to the subsequent hybridization of the "half2" probe conjugated to a Raman reporter. In analogy with the one-step assay, two configurations are available, having the reporter at the 5' (half2 R-5', close to the surface) or 3' (half2 R-3', far from the surface) terminus of the half2 probe (Figure 1b). Rhodamine 6G (R6G) and Cyanine 3 (Cy3) were identified as a convenient set of commercial reporters allowing comparable sensitivities thanks to their possible excitation under electronic resonance conditions around the laser wavelength used in the experiments (optical absorption occurs at $\lambda_{max}$ = 520 and 530 nm for Cy3 and R6G, respectively).

In silico calculations were performed to obtain an estimation of the oligonucleotide length and thus of the distance of the reporter molecule from the surface The maximum spacing was approximately 7.5 nm for the miRNA R-5' and half2 R-3' configurations, while the lowest one, calculated for miRNA R-3', was found below 1 nm. Instead, an intermediate distance of 3.5 nm was considered for the half2-5' configuration. All these calculations assumed that the hybridized complex was oriented perpendicularly to the NP surface. So that, this is a simplified model of DNA-particle and DNA-DNA interactions, as it is known that thiolated oligonucleotide monolayers can be arranged in tilted strands.[20] A slightly shorter distance can be therefore expected. However, big tilt angle are unlikely in the studied system, as it was previously demonstrated that the SERS signal of the half2 probe and of probe-222 was almost identical[6], showing that their SERS spectra are dominated by the vibrational pattern of nucleobases belonging to the first part of the two sequences. This suggest no significant tilting is occurring, as more nucleobases of the probe 222 would contribute to SERS pattern in case of big tilt angles, due to the reduced distance from the surface. Further complexity is given by the nanostructured surface, compared to a flat metallic one.

**One-step assay: miRNA R-5' vs. miRNA R-3'**

To compare the SERS performance of the different configurations, the Ag-PSD substrates were then incubated according to the two functionalization protocols with the different miRNA R-3'/5' and half2 R-3'/5' sequences, varying the miR-222 concentration in the 100 nM-0.1 nM range. The average spectra of SERS maps obtained with the miR-222 R6G-3'/5' are reported in Figure 2. An intense R6G vibrational pattern appears in the spectra of the samples incubated with a 100 nM miRNA concentration, overcoming any mode of the probe-222 down to 10 nM concentration. The typical C-C stretching of the xanthene ring at 1364 and 1646 cm$^{-1}$[21] are clearly observed, together with all the characteristic bands of R6G, as assigned in Table S1. It is worth mentioning the isolated band around 645 cm$^{-1}$, attributed to the C-C ring in plane bending of the xanthene ring[22] modified with an additional carboxylic group, due to its remarkable intensity and negligible superposition with the DNA probe bands. For this reason, such mode was selected to track the signal of all the R6G-labelled oligonucleotides, also for quantitative purposes. If the high concentration spectra of the two configurations are compared, the miRNA R6G-3' SERS signal clearly exhibits a greater intensity with respect to the configuration having the reporter far from the NPs surface. Indeed, an enhancement of more than 3 times of the fluorophore signal is detected, explained by the increased electromagnetic field intensity due to the reduced distance from the NPs surface. Actually, some experimental works investigated the distance-dependence of SERS intensity. The greatest drop of the plasmonic field intensity was shown to occur in the first nanometres, while for an increased spacing a slower decay was observed. The fastest decay of the SERS intensity was observed in a high spacing resolution study by Masango et al. that reported a signal decrease to the 20% of the "in contact" one at 0.7 nm from the plasmonic surface.[23] Different experimental works agree instead on a SERS intensity decrease to the 7-10% of the zero-spacing signal at around 3-3.5 nm[23,24] and on its further five-fold decrease when the distance from the enhancing metal surface is raised from ~0.9 to ~7 nm[25,26]. However, a great variability is documented in such literature reports, especially concerning the slope of the SERS signal enhancement decrease in close proximity of the surface. Indeed, the derived distance-dependence of SERS enhancement may vary due to several factors, including the morphology of the employed SERS substrate and the nature of the spacing layer. Despite this, based on the available literature and taking into account a slight degree of uncertainty related to the orientation and position assumed by the reporter, the three-fold enhancement observed for the miRNA R6G-3' configuration seems to be a reasonable result.

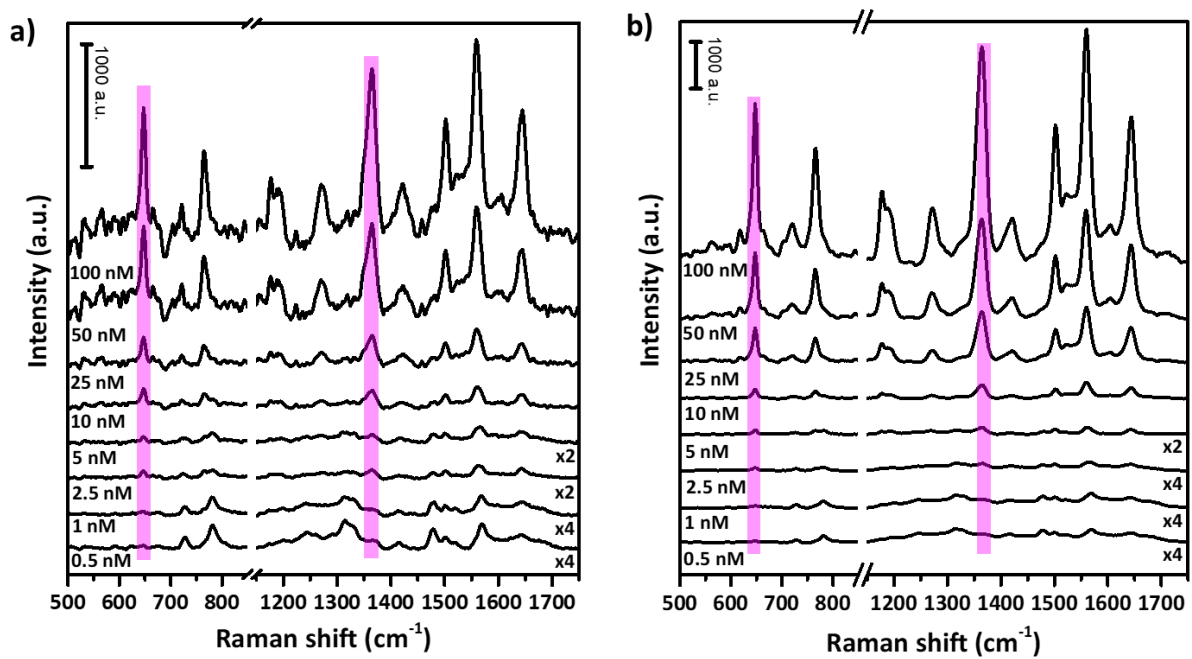

Figure 2. Average SERS spectra obtained from maps on the Ag-PSD substrate functionalized with the probe-222 according to the one-step hybridization protocol and incubated with different concentrations of a) miR-222 R6G-5' and b) miR-222 R6G-3'. The violet bars highlight two main bands of R6G (647 and 1364 cm$^{-1}$). When indicated on the right, the spectra were multiplied by the specified factor.

Similar conclusions can be drawn for the Cy3 reporter (SERS spectra at different miRNA concentrations can be found in Figure S2), although the SERS intensity of the two configurations seems to differ in most of the cases by a factor higher than 4. The described behaviour can be appreciated by monitoring one of the main SERS bands of the indocarbocyanine molecule, the vibrational mode located at around 1463 cm$^{-1}$, attributed to the $CH_3$ deformation of the ring substituents[27] (see Table S1 for a more complete assignment of the vibrational pattern of Cy3). It should be finally noted that the distance of the reporter from the metal surface also affects the typical fluorescence of the dye. The raw SERS spectra of the 5' configurations are indeed characterized by a strong fluorescence background, yielding a greater noise compared to the 3' one (Figure S3). Such fluorescence is significantly quenched when the reporter is close to the metal surface. A contribution of the improved signal-to-noise ratio to the whole SERS intensity increase for the 3' configuration cannot thus be excluded.

As a general trend, it can be further noticed that the difference in terms of SERS intensity between the close and far-to-the-surface configurations tends to decrease by lowering the miR-222 concentration, especially below 10 nM (the SERS spectra of selected concentrations of miRNA-R6G-5' and miRNA R6G-3' are superimposed in Figure S4), sometimes fading for the lowest concentrations. The comparison between the calibration curves obtained for the 3' and 5' configurations reported in Figure 3 clearly highlight such a phenomenon. This outcome is also reflected into the calculated Limits of Detection (LODs) reported in Table 1. Surprisingly, the

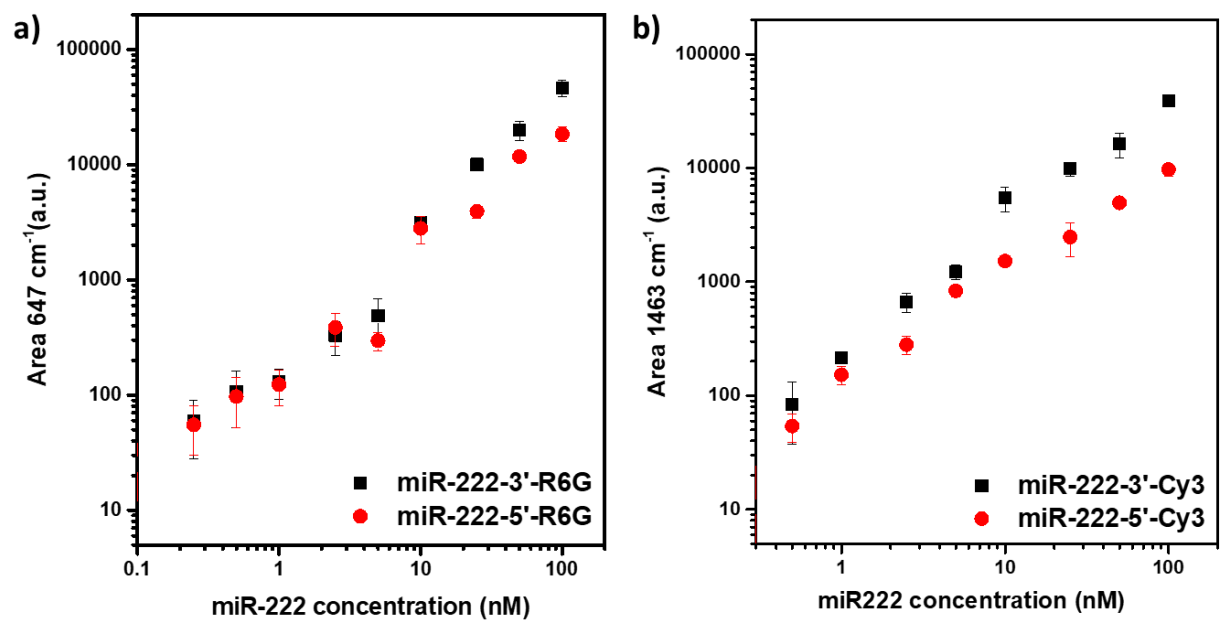

Figure 3. Average area of the main vibrational band of the miR-222 R-5'/3' vs. miR-222 concentration obtained by SERS mapping of the Ag-PSD substrates functionalized according to the one-step assay. a) R6G and b) Cy3 were employed as Raman reporters. Black squares represent the close-to-the-surface (3') configuration, while red circles the far-from-the-surface (5') one. The error bars correspond to the standard deviations.

LOD appears comparable for the Cy3-labelled miRNAs and even slightly lower for the miR-222 R6G-5' compared to the miR-222 R6G-3'. Such unexpected result depends however on the range of concentrations selected for the calculation. Indeed, a linear regime was identified below 25 nM, so that for most of the calibration curves the linear regression was performed between 10 nM and 0 nM, including several low concentrations data points.

Table 1. Limit of detection for the one-step assay performed by exploiting the four different configurations along with the miR-222 concentrations used for the linear regression.

|  | R6G | | Cy3 | |
| --- | --- | --- | --- | --- |
|  | 3' | 5' | 3' | 5' |
| Linear range (nM) | 0 - 5 | 0 – 2.5 | 0 - 5 | 0 - 10 |
| $R^2$ | 0.974 | 0.961 | 0.980 | 0.978 |
| LOD (pM) | 346 | 324 | 285 | 301 |

These results suggest a different enhancement mechanism only for the low concentrations, inducing the levelling-off of the difference between the two configurations.

Most likely, different factors could contribute to the observed trends. First, an influence of the steric hindrance for the miRNA labelled at the 3' terminal is expected. Such effect could be particularly detrimental in the case of low concentrations approaching the values of the dissociation constants between probe and miRNA, which are expected to be in the nM range[28]. Moreover, a second and probably dominant contribution can be ascribed to the SERS-active nanostructures themselves. In fact, at high miR-222 concentration it is expected that all the NPs are able to provide a sufficient electromagnetic field amplification, through the excitation of their Localized Surface Plasmon Resonances (LSPRs), to observe a SERS signal from the quite large population of miRNA-probe hybrids probed by the laser beam. In contrast, if few miRNA molecules are captured from the incubation solution, only the most efficient Raman hot-spots will be able to provide the needed enhancement.

**Electromagnetic near field analysis for Ag-PSD nanostructures**

FEM simulations were performed to corroborate such hypothesis by analysing the electric near-field (NF) intensity profiles inside and outside the gap of a dimer of Ag hemispheres in contact with the pSi surface, which was selected as simplified model of the plasmonic nanostructures. Based on the morphological parameters extracted from the FESEM analyses of the Ag-PSD substrates, different NPs diameters (25-30-35), centred at the average one, were considered for the simulations. For each diameter, gap values were varied between 2 nm and 8 nm, with a 2 nm step. Very small gaps (< 2 nm) were excluded from the studied range, as they can hardly accommodate the probe-miRNA hybrid, even if horizontally adsorbed on the NP surface, while 6 and 8 nm are expected to provide enough space for a successful hybridization. Representative spectra reporting the maximum electric NF intensity calculated within the gap between the Ag hemispheres are shown in Figure 4. All the spectra are featured by multiple resonances extending towards wavelengths larger than the excitation laser line when the inter-particle gap size is decreased and the particle diameter is increased, in agreement with the previous literature[29,30]. The highest NF intensity is observed for the longest wavelength resonance, located between 500 and 600 nm, depending on the size-gap combinations. Moreover, it should be noted that most of the simulated dimers present an intense band around the excitation wavelength used for the SERS measurements (514.5 nm), supporting the well-known high efficiency of the Ag-PSD substrates under green light excitation. Such resonance seems to originate from a splitting of the main one due to inter-particle coupling when the gap becomes smaller and smaller, as suggested by their partial convolution for the larger gap – smaller diameter pairs.

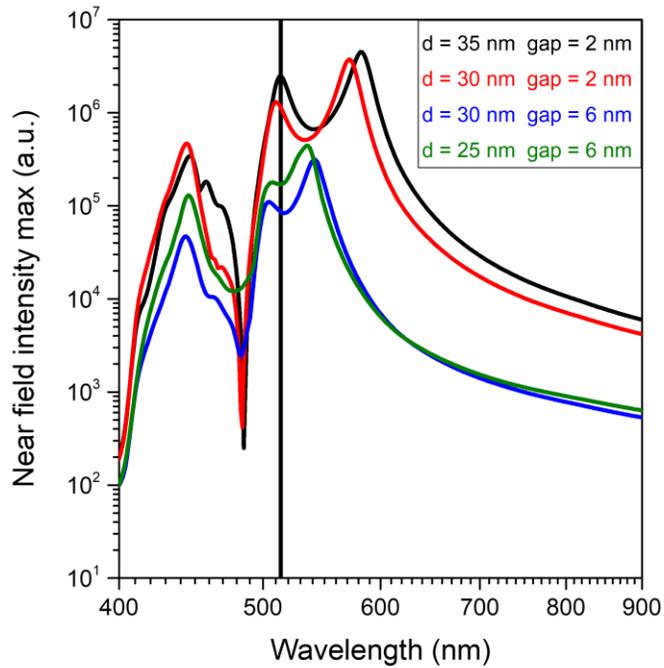

Figure 4. Maximum electric NF intensity calculated within the gap of a dimer of Ag hemispheres on a pSi surface by FEM simulations for different wavelengths and diameter-gap combinations of the AgNPs. The black vertical bar at 514.5 nm represents the excitation wavelength used for the SERS measurements.

Panel a) and b) in Figure 5 show instead the profiles of the electric NF intensity at 514.5 nm extracted at different position within and outside the gap. R = 15 nm and G = 6 nm were selected as the most representative particles radius and gap size for the investigated system. For what concerns the z profiles, shown in panel a), the electric NF decays by order of magnitudes moving from z = 0 nm (the pSi surface level) to 15 nm. Curves extracted at different x location (x = 0 corresponds to the gap centre) feature different trends only for very short distances from the pSi surface: in such range the highest intensities and the fastest decrease are observed close to the AgNP surface (x = -2, -3 nm), while lower intensities but a slower decay are detected if the profile is taken near the gap centre. The described trend is more pronounced when the gap becomes larger, with respect to the NP size (Figure S5), but, as anticipated, the differences strongly reduce at z > 2 nm for all gap-size combinations. The slight variation of the electric NF intensity along the interparticle axis beyond z = 3 nm can be better appreciated from the x profiles obtained at different z values reported in panel b) of Figure 5, where the symmetry of the NF intensity distribution with respect to the inter-particle gap center is verified.

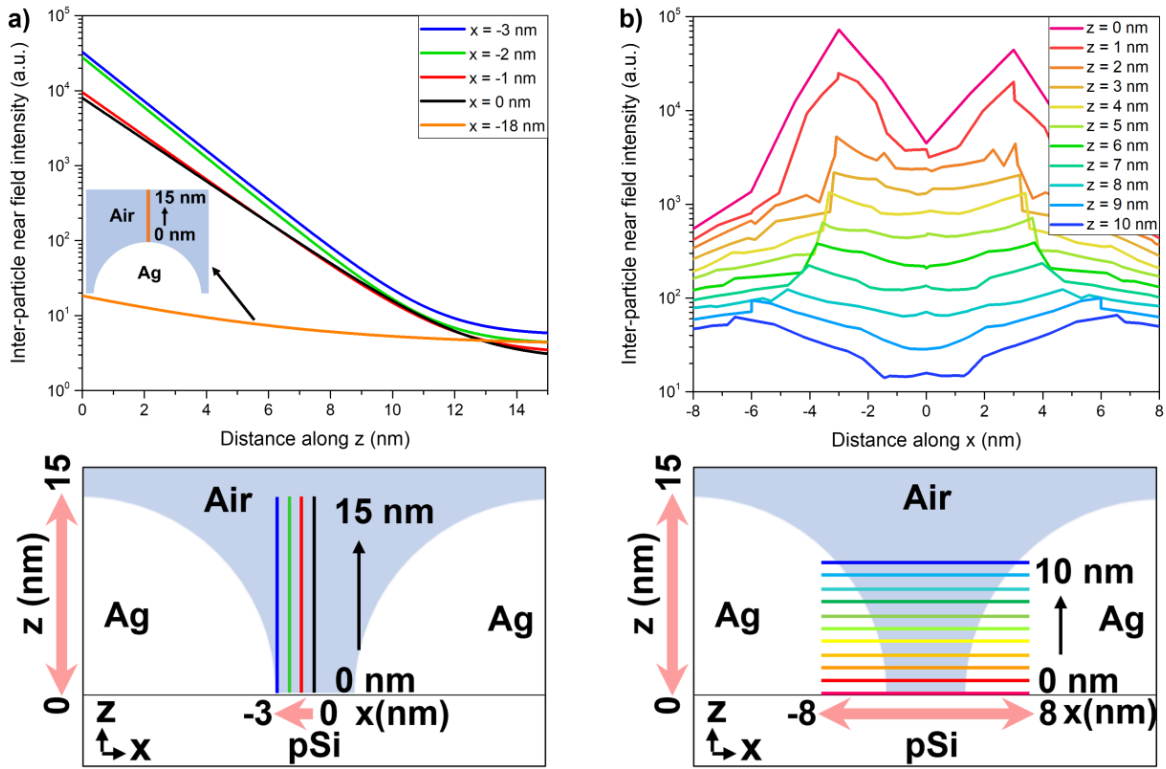

Figure 5. Spatial profiles of the electric NF intensity calculated within the gap (G = 6 nm) of a dimer of Ag hemispheres (R = 15 nm) on a pSi surface by FEM simulations a) along the z direction and b) along the x direction. The excitation wavelength was set at 514.5 nm. The inset in panel a) shows a z profile taken outside the gap (at x = -18 nm).

Some significant differences between the electric NF intensities at the gap centre and edges appear again for quite high z values (> 9 nm), when the distance between the NPs surfaces is 12 nm. An interparticle separation of the order of the particle radius is indeed considered the cut-off beyond which interparticle coupling contribution cease to dominate. The inset in panel a) of Figure 5 finally shows a z profile taken outside the gap (from the top of one of the particles, at x = -18 nm, z from 15 to 30 nm). Here, as expected, the electric NF intensity is several orders of magnitude lower than in the gap and monotonically decreases while the distance from the Ag hemisphere increases. It can be concluded that, in the case of a probe-miRNA hybrid in an interparticle gap, a Raman reporter is expected to benefit from a different enhancement depending on its position along the probe only if the oligo is bound to the NP very close to the Ag/pSi interface. Instead for $z \geq 3$ nm, the most likely situation, the electric NF experienced by the dye molecule is nearly constant throughout the gap at fixed z. At higher z values, some limited differences may however arise due to the tilted orientation of the hybrids in an upright conformation with respect to the pSi surface. Outside the gap, instead, the SERS enhancement at 7.5 nm from the NP surface is reduced to about the 15% of the one observed at 1 nm spacing. The scheme in Figure 6, representing miRNA molecules hybridized to the probe222 in or outside the gap between two AgNPs on a pSi surface, highlights the difference in the experienced NF intensity that characterizes the reporter in the 5' and 3' configurations outside the gap. However, as the electric field distribution is rather homogenous in the gap between the nanoparticles, the beneficial effects of decreasing the reporter-to-surface distance are strongly reduced or even cancelled depending on the specific combination of NP diameter, gap size and binding site of the probe. It should be in fact underlined that, even if limited, the heterogeneity in the morphology of the SERS substrate can obviously affect the enhancement of the SERS signal obtained when the reporter at the 5'- terminus of the miRNA is located in the gap. Indeed, the reporter can be even in contact with the neighbouring NP in some cases. Unfortunately, the prevalence of the gap contribution to the SERS signal at low miRNA concentration does not allow the expected improvement of the LODs moving from the far- to the close-to-the-surface reporter configuration. Though, it should be noted that at the same time, the signal is already boosted in the gap for the miRNA R-5' configuration at low concentrations. As a result, the slope of the calibration curve is reduced compared to the high concentration regime.

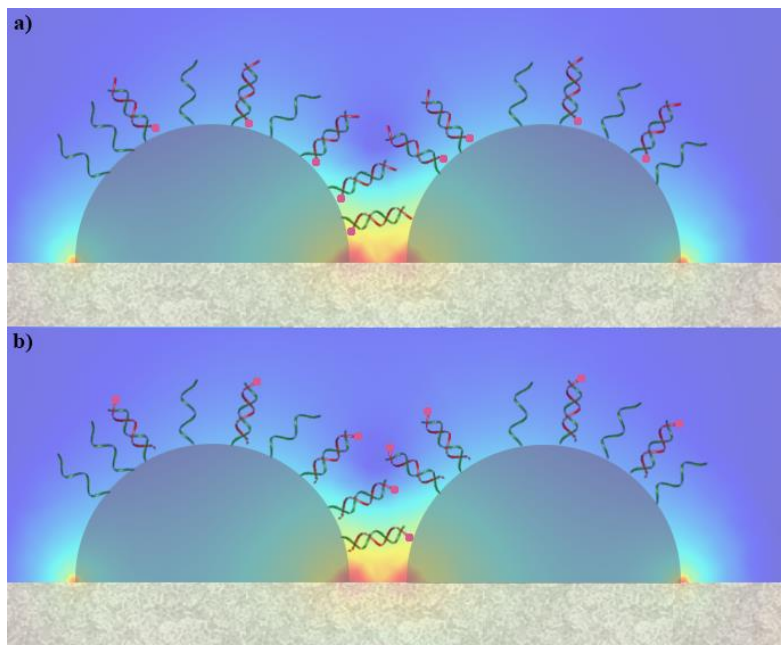

Figure 6. Scheme representing miRNA molecules hybridized to the probe-222 in or outside the gap between two AgNPs on a porous silicon surface for both the a) 3' and b) 5' configurations of the Raman reporters.

A further indication of a distinct origin of the SERS enhancement for the high and low concentrations of target miRNA is indeed provided by the non-linear trend often observed in SERS calibrations. In fact, the sensitivity should change if all the particles or only the hot-spots contribute to the SERS intensity. Modelling this variation is a great challenge, due to the strong dependence of such effects on the SERS substrate morphology. Nevertheless, some attempts can be found in the literature that use non-linear fitting functions, such as the Langmuir or dose-response curves[31,32]. Actually, none of the two models provided a satisfactory fit of the current data, showing that adsorption models are not always adequate to describe the trend of the SERS intensity vs. the analyte concentration, probably because they do not take into account such hot-spot related effect that becomes evident only for very low concentrations of the target analyte. It should be mentioned that similar non-linearities were in many cases attributed to the SERS Intensity Fluctuations due to single molecule events in the nanomolar and subnanomolar concentration range.[1] Namely, it is assumed that such fluctuations depend on the motion of the analyte in and outside the available hot-spot, whose volume sharply decreases for increasing enhancement.[3] Though, such phenomena shouldn't be the source of non-linearity in the presented system, since the functionalization of the surface does not allow great motion of the involved molecules and the SERS measurements are not acquired in solution, where some dynamic equilibrium between bound and free miRNAR5'/3' could establish.

**Two-step assay: half2 R-3' vs. half2 R-5'**

Different concentrations of unlabelled miR-222 were incubated on the half1 functionalized Ag-PSD substrates and afterwards detected through the hybridization with the half2 probe labelled with the different reporters both at the 3' or 5' terminus. The results of the SERS maps acquired on the described samples are displayed in Figure 7 for the 5' and 3' R6G-labelled half2 probe. As expected due to the reduced difference in the reporter-to-surface distance (3.5 nm vs. 7.5 instead of < 1 nm vs. 7.5 in the case of the labelled miR-222) in the two configurations the SERS intensity was only moderately enhanced when the reporter was located close to the AgNPs (generally around or less than 2 times the area of the Raman band of R6G at 647 cm$^{-1}$ calculated for the 3' labelling).

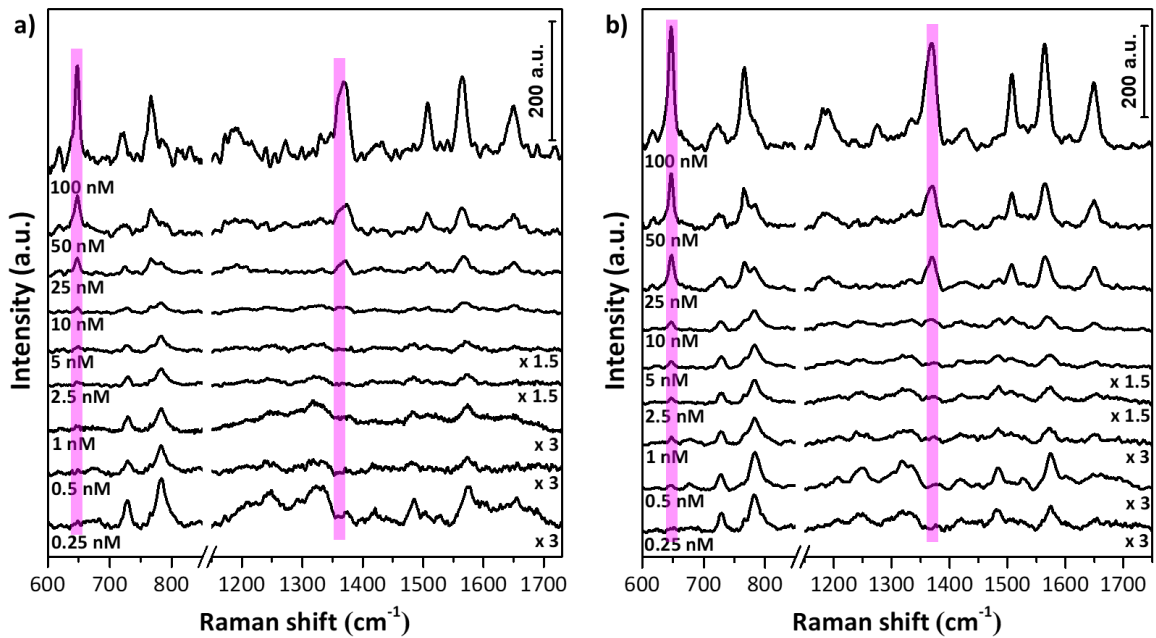

Figure 7. Average SERS spectra obtained from maps on the Ag-PSD substrate functionalized with half1, exposed to different concentrations of miR-222 and then incubated with a) half2 R6G-3' and b) half2 R6G-5', according to the two-step hybridization protocol. The violet bars highlight the two main bands of R6G (at 647 and 1364 cm$^{-1}$). When indicated on the right, the spectra were multiplied by the specified factor.

Differently from the case of the one-step assay, such enhancement does not fade for concentrations around and lower than 5 nM. A close inspection of the superimposed average SERS spectra (Figure S6 and S7) highlights that some differences between the two configurations are preserved in most cases also in the nanomolar range for both the R6G and Cy3 reporters. Such trend is also confirmed by the calibration curves of SERS area vs. miR-222 concentrations reported in Figure 8, that were derived considering the integrated area of the same SERS bands analysed for the one-step assay. This outcome is reflected in the calculated LODs, which result 1506 pM and 357 pM when the half2 R6G-3' and the half2 R6G-5' are concerned, respectively. An improvement of the sensitivity of the assay is thus achieved for the two-step assay by moving the reporter towards the nanoparticle surface, despite the order of magnitude of the LOD doesn't change. It should be noted that a similar benefit is obtained using the half2-Cy3, for which the LODs are 988 nM and 627 pM, as listed in Table 2.

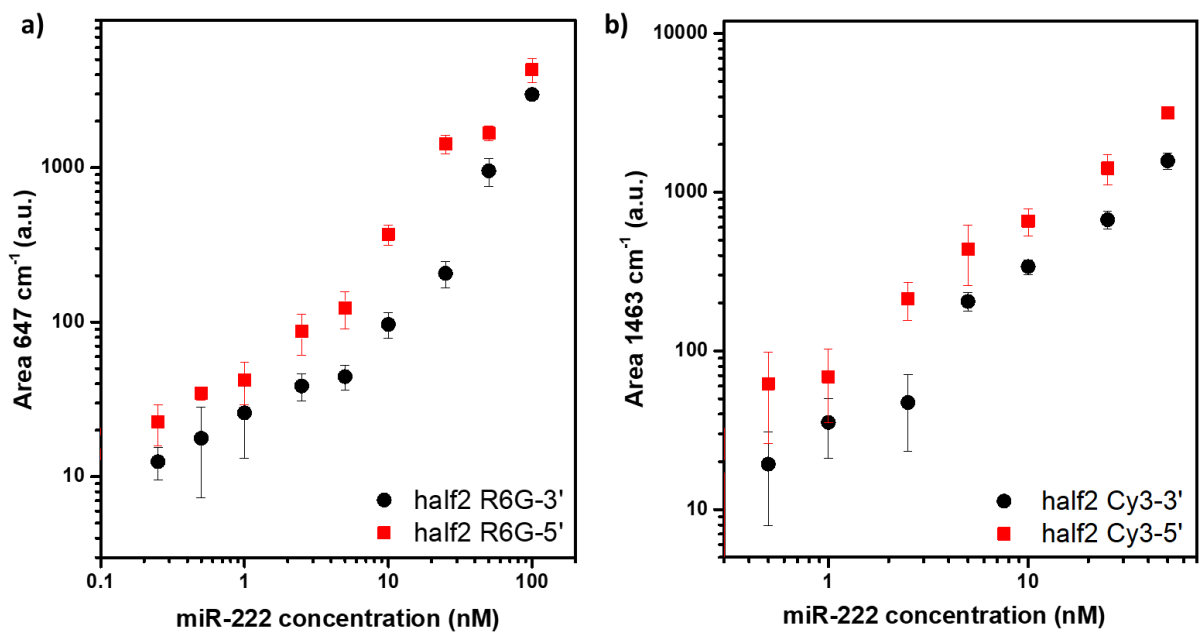

Figure 8. Average area of the main vibrational band of the half2 R-5'/3' vs. miR-222 concentration obtained by SERS mapping of the Ag-PSD substrates functionalized according to the two-step assay. a) R6G and b) Cy3 were employed as Raman reporters. Red squares represent the close-to-the-surface (5') configuration, while black circles the far-from-the-surface (3') one. The error bars correspond to the standard deviations.

Table 2. Limit of detection for the two-step assay performed by exploiting the four different configurations along with the miR-222 concentrations used for the linear regression.

|  | R6G | | Cy3 | |
|---|---|---|---|---|
|  | 3' | 5' | 3' | 5' |
| Linear range (nM) | 0 - 10 | 0 – 10 | 0 - 10 | 0 - 10 |
| $R^2$ | 0.943 | 0.977 | 0.952 | 0.979 |
| LOD (pM) | 1506 | 357 | 988 | 627 |

Actually, the outcomes appear to be different from the one detected in the one-step assay, as confirmed by the repetition of the experiments. Such discrepancies cannot however depend on a diverse electromagnetic enhancement mechanism: hot-spot contribution should again dominate at low miR-222 concentration. It can be therefore hypothesized that they depend on the impact of the position of probe/miRNA labelling on the hybridisation efficiency. Indeed, it is well known that steric hindrance effects play a significant role in duplex formation when the hybridization involves a surface-immobilized probe, such as in microarrays.[33,34] As an example, an excessive packing of probes at the surface is detrimental, since it results in slow hybridization kinetics and lowered degree of probe/target hybridization.[33,35] Moreover, it was observed that the hybridisation yield increases when probe-target binding occurs farther from the support surface, due to an improved accessibility of the oligonucleotide sequence[36] For this reason spacer linkers are often employed to increase the distance of the strand and reduce steric interference from the support.[36] In such a framework, the presence of a bulky reporter at the close-to-the surface terminus of the half2 probe or miR-222 is expected to enhance steric hindrance related issues with respect to the far-from the surface configurations. However, if the one step and two step assay are compared, it is immediately apparent that the half2 R-5'/miR-222 hybridization occurs in a less sterically hindered environment than the miRNA R-3'/probe 222 heteroduplex formation. Indeed, in the two step assay, the second hybridization takes place at an increased distance from the NP surface. It is therefore reasonable that the observed trends arise from the combination of electromagnetic and hybridisation efficiency effects: the electromagnetic near field intensity dependence on the reporter-to-surface distance strongly reduces in the low concentration regime for all the tested configurations, but the higher LODs for the miR-222 R-3' configuration compared to the miR-222 R-5' one are probably ascribable to the severe steric hindrance effects in the first case. It should be underlined, that the influence of steric constraints would probably not be appreciated in the absence of the levelling off of the electromagnetic enhancement. Instead, for the two-step assay, where steric issues are relaxed for the second hybridization step (the one that involves the reporter), a slightly higher SERS intensity is still observed at low miR-222 concentration for the close-to-the surface configuration compared to the 3' one. It should be noted, indeed, that the inherent lower melting temperature of the half1/miR-222 and miR-222/half2 complexes compared to the longer and thus more stable probe222-miR-222 duplex was previously compensated by the optimization of the hybridization conditions[7].

Despite lower than initially expected, due to the described increased contribution of hot-spots to the SERS signal observed at low target concentrations, an increase of the sensitivity of the two-step protocol for miRNA detection is attained by reducing the distance between the reporter and the plasmonic NPs. Such improvement can be fruitfully exploited to enhance the detection of such biomarkers in real samples, due to the label-free approach featuring the assay.

**Conclusions**

Ag-porous silicon-PDMS SERS substrates were exploited to investigate the role of the Raman reporter position along the oligonucleotide sequence in the framework of a one-step and a two-step hybridization assay for the SERS detection of miRNAs. As for the one-step assay, a significant variation (more than 6.5 nm) of the reporter-to-surface distance could be obtained by labelling the miRNA at the 3' and 5' terminus. The SERS analysis of several miR-222 R-5'/3' concentrations showed a clearly greater SERS intensity for the 3' configuration compared to the 5' one over 5 nM; such improved SERS detection gradually vanished at lower miRNA concentrations. FEM simulations of the electric near field for a model Ag dimer on a porous silicon surface suggested that the unexpected concentration-dependent outcome could be compatible with an increased hot-spot contribution to the whole SERS signal in the nanomolar range of miR-222 concentrations. On the other side, for the two-step assay, a change in the reporter-to-surface spacing of about 4 nm was possible, by labelling the second half2 probe at the 3' or 5' terminus. The reduced variation of the reporter-to-surface distance between the two detection configurations yielded a mitigated difference between the SERS intensity observed for the two labelling positions. Interestingly, the LOD for miR-222 was always lowered for the close-to-the-surface configuration, in contrast to the case of the one-step assay. Such result, that was confirmed using two different Raman reporters (R6G and Cy3), points out the possible role of steric hindrance in decreasing the hybridization efficiency in the case of the 3' labelled miRNA. On the whole, the study allowed to improve the sensitivity of the two-step assay and provided new insight on the complex balance between plasmonic effects and surface chemistry in the definition of a SERS bioassay performance.

**Supporting information**

FESEM characterization of Ag-PSD substrates, assignments of the main SERS bands of reporters, SERS analyses of different miR-222 concentrations using Cy3 as reporter, electric NF within the gap of Ag dimer with varying geometrical parameters, typical fluorescence background in the raw SERS of the 5' and 3' configurations.

**Acknowledgements**

Financial support from the POR FESR 2014-2020 Piedmont Regional Projects "Digital Technology For Lung Cancer Treatment"—DEFLECT (2018–2022) is gratefully acknowledged. We gratefully acknowledge Dexmet Corporation for providing the copper grids used for the FESEM imaging.

**Label-free detection of miRNA: role of probe design and bioassay configuration in Surface Enhanced Raman Scattering based biosensors**

*Chiara Novara[1], Daniel Montesi[1], Sofia Bertone[1], Niccolò Paccotti[1], Francesco Geobaldo[1], Marwan Channab[1,2], Angelo Angelini[2], Paola Rivolo[1], Fabrizio Giorgis[1], Alessandro Chiadò[1]*

[1]*Department of Applied Science and Technology, Politecnico di Torino, Corso Duca degli Abruzzi, 24, 10129 Turin, Italy.*

[2]*Advanced Materials and Life Sciences, Istituto Nazionale di Ricerca Metrologica (INRiM), Strada delle Cacce 91, Turin 10135, Italy.*

**SERS substrate morphology**

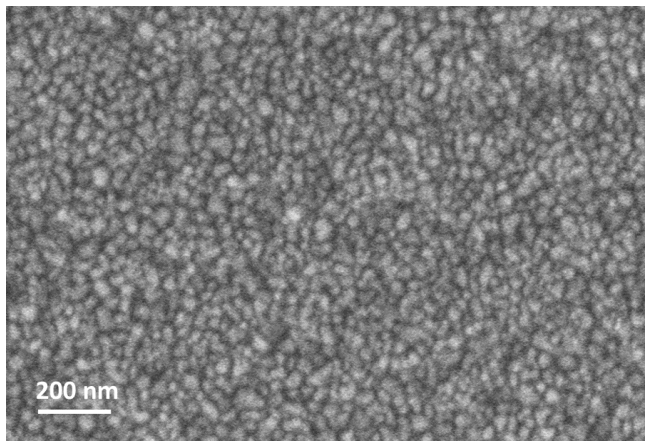

**Figure S1**. FESEM top view of an Ag-coated porous silicon membrane.

**One-step assay: miRNA R-5' vs. miRNA R-3'**

| R6G | | Cy3 | |
|---|---|---|---|
| Raman shift (cm$^{-1}$) | Assignment | Raman shift (cm$^{-1}$) | Assignment |
| 647 | C–C ring in-plane bending in xanthene/phenyl rings[1,2] | 1170 1190 | C–H in-plane bending in N-alkyl substituents[3] |
| 765 | C–H out-of-plane bending[1,2] | 1263 | ring stretching[3] |
| 1127 | C–H in-plane bending in xanthene/phenyl rings[1,2] | 1388 | Methine chain[3] |
| 1178 | C–H plane bending in xanthene ring[2] | 1401 | CH$_3$ symmetric deformation[3] |
| 1270 | hybrid mode (xanthene/ phenyl rings)[2] | 1462 | CH$_3$ asymmetric deformation[3] |
| 1364 | C–C stretching in xanthene ring[2] | 1583 | C=N stretching[3] |
| 1502 | C–C stretching in xanthene ring[2] | | |
| 1558 | C–C stretching in phenyl ring[2] | | |
| 1644 | C–C stretching in xanthene ring[2] | | |

**Table S1.** Assignments of the main vibrational bands of R6G and Cy3 reporters.

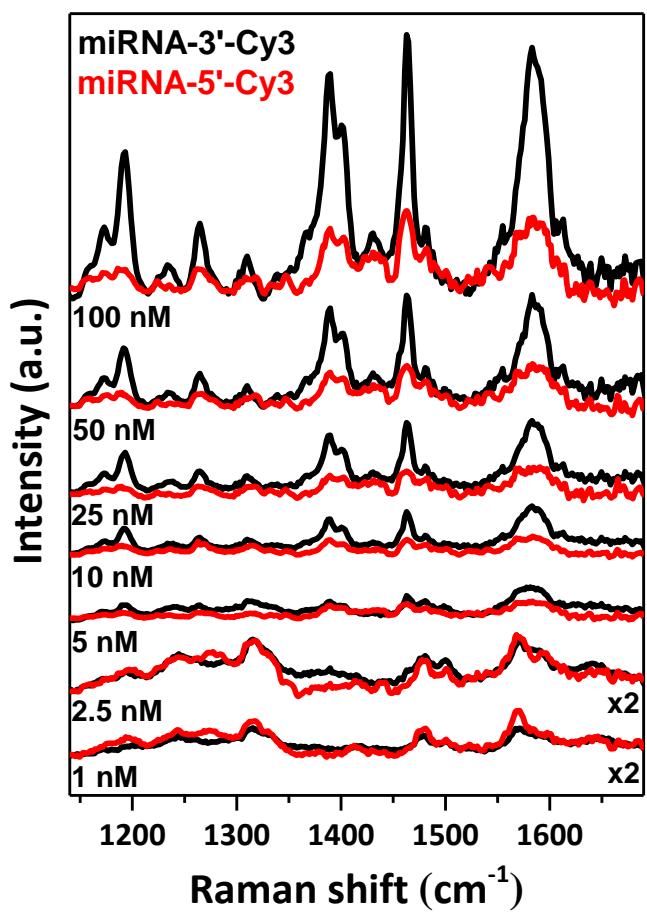

**Figure S2**. Average SERS spectra obtained from maps on the Ag-PSD substrate functionalized with probe-222 according to the one-step hybridization protocol and incubated with different concentrations of miR-222 Cy3 5'(red curves)/3'(black curves). The coloured bars highlight the main band of Cy3 at around 1463 cm$^{-1}$. When indicated on the right, the spectra were multiplied by the specified factor.

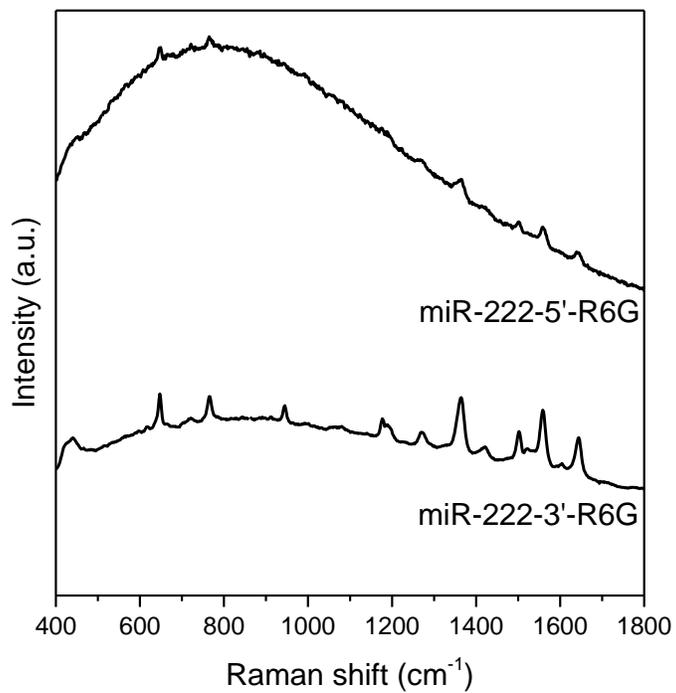

**Figure S3**. Comparison of fluorescence background of raw SERS spectra acquired on the Ag-PSD substrates functionalized according to the one-step hybridization protocol and incubated with miR-222 R6G 5'- (top curve) and miR-222 R6G-3'- (bottom curve).

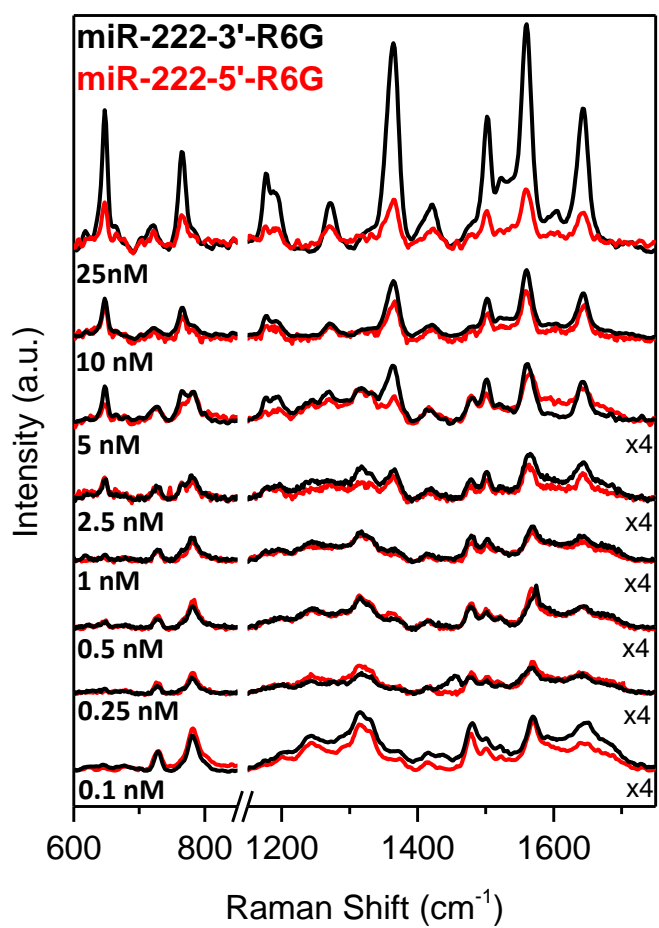

**Figure S4**. Direct comparison of selected average SERS spectra obtained from maps on the Ag-PSD substrate functionalized with probe-222 according to the one-step hybridization protocol and incubated with different concentrations of miR-222 R6G-5'- (red curves) and miR-222 R6G 3' (black curves). When indicated on the right, the spectra were multiplied by the specified factor.

**Electromagnetic near field analysis for Ag-PSD nanostructures**

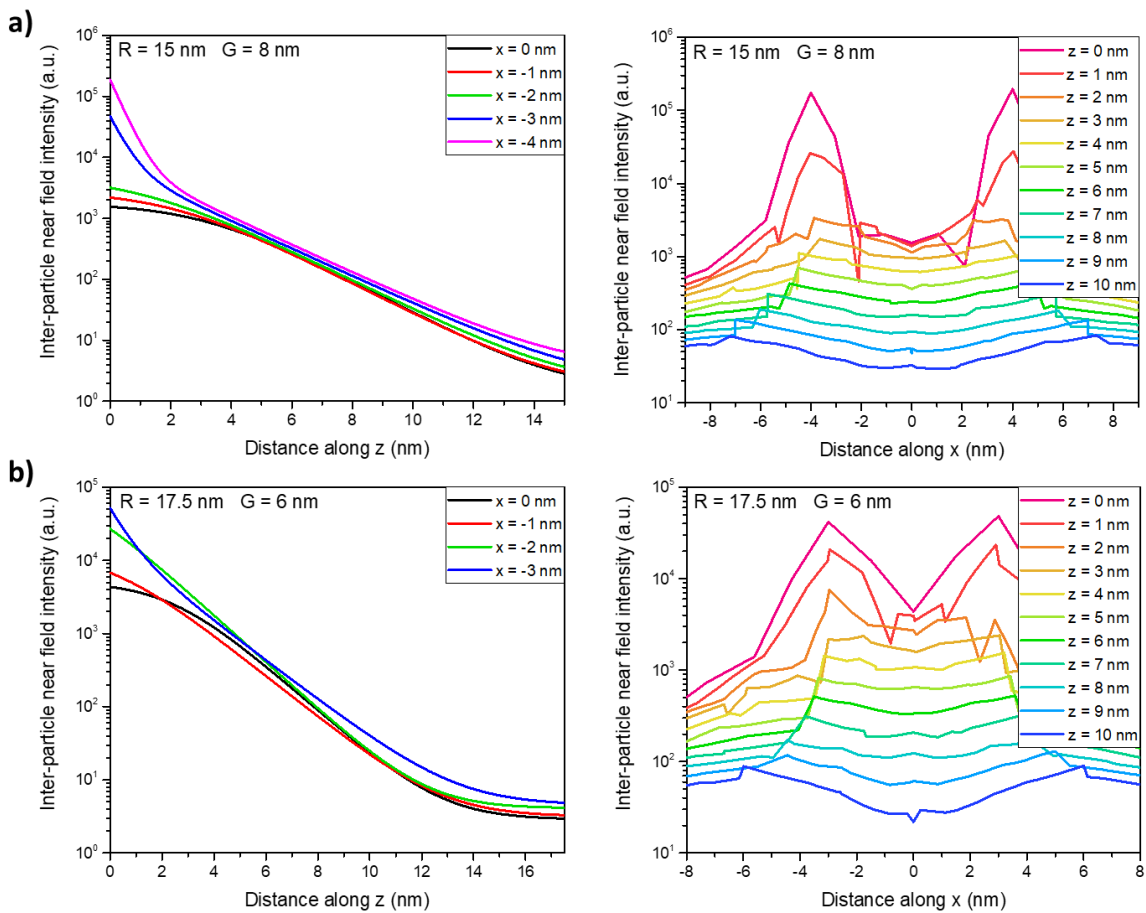

**Figure S5.** Spatial profiles of the electric NF intensity calculated within the gap of a dimer of Ag hemispheres on a porous silicon surface by FEM simulations along the z direction (left) and along the x direction (right). a) interparticle gap (G) = 8 nm, particle radius (R) = 15 nm: b) interparticle gap (G) = 68 nm, particle radius (R) = 17.5 nm The excitation wavelength was set at 514.5 nm.

**Two-step assay: half2 R-3' vs. half2 R-5'**

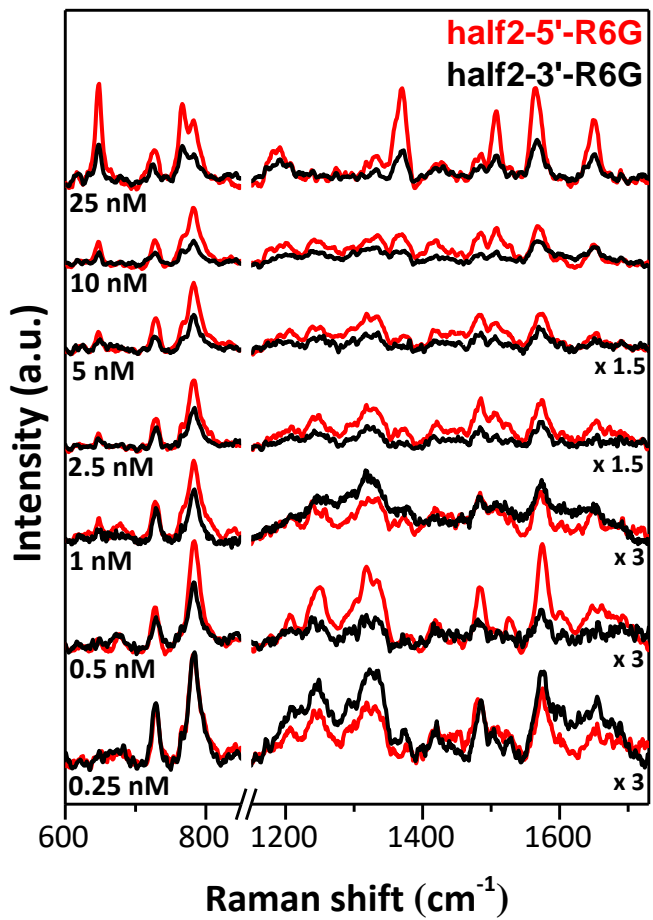

**Figure S6.** Direct comparison of selected average SERS spectra obtained from maps on the Ag-PSD substrate functionalized with half1, exposed to different concentrations of miR-222 and then incubated with half2 R6G-5' (red curves) and half2 R6G-3'(black curves), according to the two-step hybridization protocol. When indicated on the right, the spectra were multiplied by the specified factor.

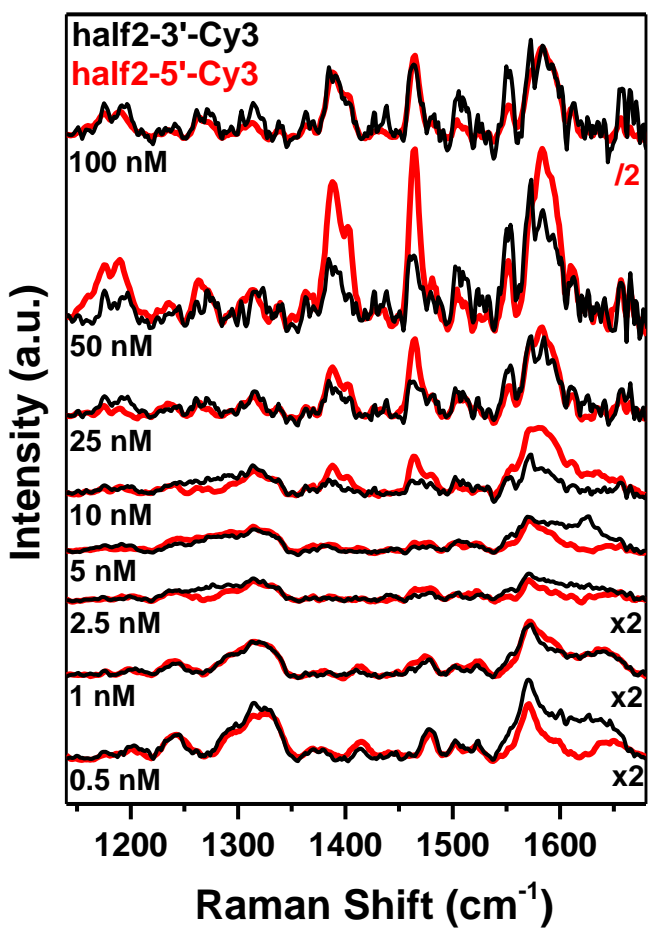

**Figure S7.** Direct comparison of selected average SERS spectra obtained from maps on the Ag-PSD substrate functionalized with half1, exposed to different concentrations of miR-222 and then incubated with half2 Cy3-5' (red curves) and half2 Cy3-3' (black curves), according to the two-step hybridization protocol. When indicated on the right, the spectra were multiplied by the specified factor.